\documentclass[journal,10pt]{IEEEtran}
\usepackage{float}
\usepackage{comment}
\usepackage[table]{xcolor}
\usepackage{cite}      
\usepackage{graphicx}  
\usepackage{psfrag}    
\usepackage{placeins}
\usepackage{url}
\usepackage{subcaption}
\usepackage{caption}
\usepackage{times}
\usepackage{textcomp}
\usepackage{amsmath} 
\usepackage{amsfonts,amsthm,amssymb,mathtools}
\usepackage{balance}
\usepackage[normalem]{ulem}
\usepackage{xcolor}

\usepackage{cuted}
\usepackage{amsmath,bm}
\usepackage{hyperref}
\usepackage{cleveref}

\begin{document}
\title{Estimation of Electrical Characteristics of Inhomogeneous Walls Using Generative Adversarial Networks}
\author{Kainat~Yasmeen,~\IEEEmembership{Student Member, ~IEEE} and 
    Shobha~Sundar~Ram,~\IEEEmembership{Senior Member,~IEEE}
\thanks{The authors are with Indraprastha Institute of Information Technology Delhi (email: kainaty@iiitd.ac.in; shobha@iiitd.ac.in).}}
\maketitle

\begin{abstract}
Through-wall radars are researched and developed for the detection, localization, and tracking of human activities in indoor environments. Electromagnetic wave propagation through walls introduces refraction, attenuation, multipath, and ghost targets in the radar signatures. Estimation of wall characteristics (dielectric profile and thickness) can enable wall effects to be deconvolved from through-wall radar signatures. We propose to use generative adversarial networks (GAN) to estimate wall characteristics from narrowband scattered electric fields on the same side of the wall as the transmitter. We demonstrate that the GANs, consisting of two neural networks configured in an adversarial manner, are capable of solving the highly nonlinear regression problem with limited training data to estimate the dielectric profile and thickness of actual walls up to 95\% accuracy based on training with simulated data generated from full-wave solvers.
\end{abstract}
\providecommand{\keywords}[1]{\textbf{\textit{Keywords--}}#1}

\section{Introduction}
\label{sec:Introduction}
Through-wall and indoor radars have been widely researched for human activity detection and recognition for various applications such as security and surveillance, search and rescue, and assisted living, and fall detection \cite{amin2014through,amin2016radar}. Based on the type of radar, radar signatures could be micro-Doppler spectrograms obtained from joint time-frequency representations of time-domain narrowband radar data \cite{ram2008doppler,ram2010simulation}; high range resolution profiles generated from broadband data \cite{ram2009simulation}; range-Doppler ambiguity diagrams generated from stepped-frequency continuous wave radars or frequency modulated continuous wave radars \cite{qi2016detection} or range-azimuth signatures obtained from noise radars \cite{narayanan2008through,lai2010ultrawideband}. The advantages of the narrowband radar are that these radars can be implemented with low cost, custom-off-the-shelf components; are low noise, and have high Doppler resolution that enables the detection and recognition of dynamic target activities \cite{ram2007human,ram2008through}. Broadband radar data, on the other hand, when gathered from large real/synthetic antenna apertures, can be processed to localize targets in range, azimuth, and elevation \cite{ahmad2005synthetic,ahmad2008three}. However, when these radars are deployed in through-wall scenarios, the radar signals undergo complex propagation phenomenology such as attenuation, ringing, refraction, and multipath that result in ghost targets and other types of significant distortions in the radar signatures \cite{ram2010simulation}. Walls are considerably diverse - from single-layer homogeneous dielectric walls commonly found indoors to inhomogeneous/multi-layered exterior walls. Similarly, wall materials can range from wood, cement, brick, stone, mud, to cinder-block with air gaps. In radar literature, different strategies have been explored to handle wall-based distortions on radar signatures. One strategy is to remove wall distortions - particularly front face reflections and refraction - through signal processing methods based on complete or incomplete knowledge of wall characteristics (dielectric constant and thickness) \cite{solimene2009three,solimene2011twi}. This strategy is particularly effective if the wall is a single-layer homogeneous dielectric wall. More recently, sparsity-based deep learning methods have been exploited for removing wall effects without the requirement of any knowledge of wall characteristics \cite{ wang2017look,guo2017imaging,liu2019clutter,vishwakarma2020mitigation,ram2021sparsity}. However, these methods are predicated on the availability of a large volume of radar data of targets in free space and through-wall conditions. A third strategy has been to exploit the multipath introduced by through-wall propagation to improve the radar detection performance \cite{setlur2011multipath,ghorbani2021deep}. Other strategies have involved adapting hardware parameters to suppress wall effects \cite{guo2018multipath,zheng2018multilevel}. In all of these studies, the focus has been on detecting and localizing targets behind the walls. In our work, we focus on a fifth strategy. Here, we use learning algorithms to directly estimate wall characteristics - thickness and dielectric profile - in the absence of targets from the radar data. The purpose of estimation of the wall electrical characteristics is to subsequently use the model for deconvolving wall effects from through-wall radar signatures of targets. 

Estimation of wall characteristics from electromagnetic fields data is a classical electromagnetics inverse scattering (EIS) problem \cite{soldovieri2007through}. Traditionally, EIS problems have been tackled using deterministic approaches such as Born approximation method (BAM) \cite{habashy1993beyond} and back propagation (BP) \cite{chen2018computational, belkebir2005superresolution}. However, these techniques are known to provide poor accuracy especially when the region of interest consists of scatterers with high dielectric constants. Iterative microwave imaging methods such as  the Born iterative method (BIM) \cite{wang1989iterative}, the distorted Born iterative method \cite{chew1990reconstruction}, the contrast source-type inversion method \cite{song2005through} and the subspace optimization method \cite{chen2009subspace,zhong2011fft} have also been explored. These are time consuming and often not suitable for real time reconstruction. These approaches are further complicated by the requirement of heuristic tuning of parameters for optimizing the performance of the algorithms. Alternatively, stochastic approaches involving genetic algorithms or particle swarm optimization \cite{semnani2009reconstruction} have been researched. These methods are computationally complex in terms of time and memory, and the complexity further scales with the degree of inhomogeneity in the region of interest. Hence, these methods have generally been ineffective in solving problems where the walls are inhomogeneous and dispersive due to the nonlinear relationship between the wall characteristics and the scattered fields. 

Machine learning methods have been explored for a wide variety of EIS-based applications such as microwave imaging, ground penetration radar, remote sensing, and biomedical imaging \cite{chen2020review}. The challenges faced for the inverse problems using machine learning are limited data and high accuracy requirements. Artificial neural network-based methodologies have also been used to extract information about the geometric and electromagnetic properties of scatterers \cite{caorsi1999electromagnetic, rekanos2002neural}. However, most of these methods represent scatterers with only a few parameters, such as their sizes, positions, shapes, and relative permittivities, and cannot be used when there is significant inhomogeneity. 

In the past few years, convolutional and deep neural networks (DNNs) have been researched for solving highly nonlinear and ill-posed problems. DNNs are a flexible representation of high-dimensional nonlinear functions and have been explored for through-wall radar imaging \cite{cicchetti2021numerical,zheng2021human,li2020human}. The main challenge in the use of these algorithms is the requirement of large and correctly labeled training data gathered in diverse experimental settings. 
Generative adversarial networks (GAN), which have emerged from the deep learning community, consist of two networks that work in a zero-sum game. The two networks - generator and discriminator - could be either artificial neural networks or convolutional neural networks \cite{goodfellow2020generative,radford2015unsupervised,salimans2016improved}. GANs have been shown to be versatile in modeling/mapping highly nonlinear relationships between the input and output without requiring computationally expensive Markov chains \cite{creswell2018generative,hinton2006fast,goodfellow2020generative}. 
The advantage of GAN over other generative models such as Boltzmann machines is its ability to rapidly generate several realistic samples in parallel. In our problem statement, they provide the advantage of requiring small datasets during training compared to other algorithms with single networks. This is extremely important for real-world applications where they may be significant diversity in wall characteristics, and it may be practically impossible to train an algorithm for every possible wall type that may be encountered during radar deployment.

GANs have been recently applied for solving EIS methods \cite{liu2018generative}. The main objective of this work is to use GANs to estimate the permittivity profile and thickness of inhomogeneous walls based on simple radar measurements that can be easily carried out with custom off-the-shelf components. Specifically, the radar configuration consists of a continuous wave narrowband, linearly polarized source excitation, and scattered fields that are measured at multiple receiver locations on the same side as the transmitter in the absence of targets on the far side.
We examine GANs configured with both ANNs and CNNs, and with time (real-valued) and frequency domain (complex-valued) electric field data as input. Due to the challenges in gathering large volumes of training data, we use simulated data of scattered electric fields to train the GANs, which we subsequently use to estimate the characteristics (thickness and dielectric profile) of actual walls. We believe ours is the first work to estimate characteristics of real-world walls based entirely on training the learning algorithm with simulated data. We benchmark our proposed method with three classical electromagnetic techniques - BAM, BIM and BP and two popular machine learning techniques using a single neural network - the fully connected neural network and the convolutional neural network. BAM and BP utilize the integral electric field formulation to estimate the dielectric characteristics. The BIM introduces a regularization operator and iteration between the forward scattering and the inverse scattering problem formulations. We apply all these state-of-the-art techniques to the exact same simulation problem as the proposed GAN techniques to study their effectiveness.
The data and algorithms are provided to the interested reader at \url{https://essrg.iiitd.edu.in/?page_id=4355}. We demonstrate that we can correctly estimate the thickness of the walls up to 95\% and the dielectric constant profile within 97\%, showcasing the versatility of the GAN-based algorithms for EIS problems. We find that the chief limitation of the approach is when the wall material is largely transparent to source frequency resulting in very low scattered fields at the same side as the transmitter. 

Our paper is organized in the following manner. In the following section, we describe four different approaches for configuring GANs for estimating wall properties from the scattered electric field. In section.\ref{sec:Sim}, we describe the experimental setup for generating simulated data that are subsequently used for training the GANs and subsequently validating the GANs using test simulated data. Once validated, the GANs are tested on measurement data from actual walls. The experimental setup for gathering measurement data and the measurement results are presented in section.\ref{sec:Measurement}. This is followed by the conclusion in section.\ref{sec:Conclusion}.
\section{ Methodology}
\label{sec:Methodology}
The objective of our work is to estimate the thickness and electrical characteristics of walls based on their scattered electric field.
\begin{figure}[htbp]
\centering
\includegraphics[scale=0.65]{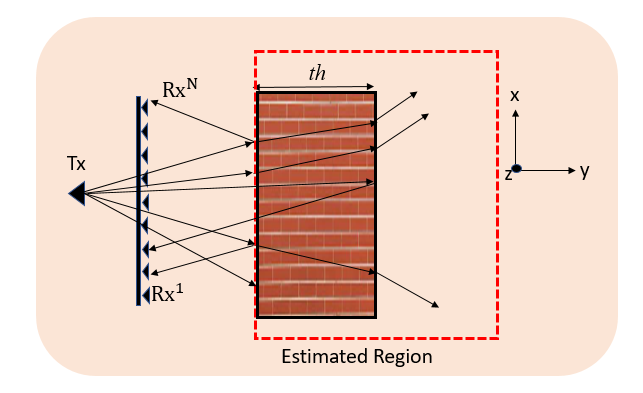}
\caption{An illustration of the electric field scattered from a wall of thickness $th$ with one transmitter (Tx) and multiple receivers ($Rx^1:Rx^N$).} 
\label{fig:scatterer}
\end{figure}
We apply the following constraints to this problem statement. First, we assume a two-dimensional (2D) Cartesian problem space as shown in Fig. \ref{fig:scatterer}, where the wall is illuminated by a narrowband vertically polarized electric field ($E_z^i$) from a single transmitter. The scattered field from the wall is measured at multiple receiver locations on the same side as the transmitter ($E_z^{n},n=1:N$). In this problem formulation, it is assumed that the walls are infinitely long and uniform along $z$ but may be inhomogeneous along $x$ and $y$ dimensions. The 2D framework is chosen to reduce the computational complexity of the problem and because most walls show homogeneity along the height. Second, we assume that there are no targets on the far side of the wall (opposite side to the transmitter), and hence the scattered field is only a function of the wall parameters; Third, we assume that the walls consist of very low conductivity materials. In other words, we assume that the dielectric constant of the walls is real. We use generative adversarial networks (GAN) to perform the inversion operation from the scattered electric field to estimate the wall thickness and dielectric profile $\epsilon_r(x,y)$. In this section, we describe how the GAN is configured to solve this problem.
\subsection{GAN Architecture}
\begin{figure*}[htbp]
\centering
\includegraphics[scale=0.6]{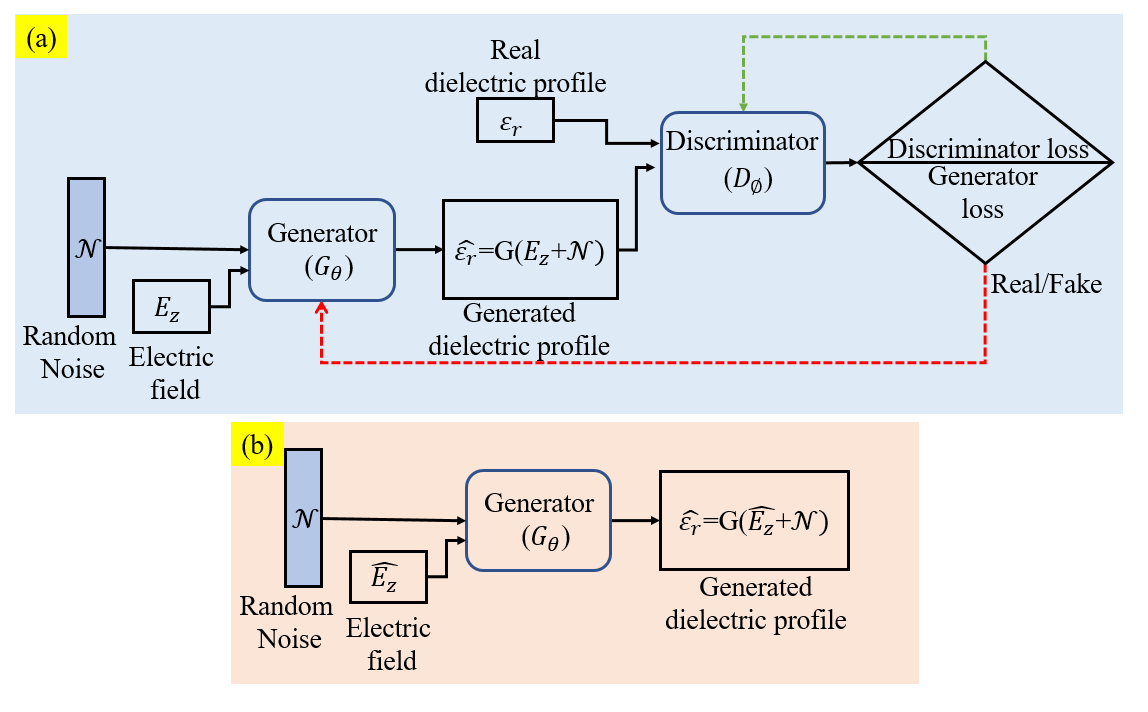}
\caption{Block diagram of GAN architecture for estimating dielectric profile from given electric field during the (a) training phase and (b) testing phase.}
\label{fig:Training}
\end{figure*}

GAN comprises two networks known as discriminator ($D_{\phi}$) with weights $\phi$ and generator ($G_{\theta}$) with weights $\theta$ as shown in the Fig.\ref{fig:Training}. 
During the training stage, the input to the generator is a latent noise vector $\mathcal{N}$ concatenated with the scattered electric field $E_{z}^n,n=1:N$ from the $N$ receiver positions. A noise vector is introduced to the input to provide randomness to the generator and prevent overfitting \cite{goodfellow2020generative}.
The electric field could be either real time-domain values vectorized to form a single column or complex frequency domain values corresponding to the source excitation frequency. In the second case, the imaginary values are concatenated to the real values to form a single column vector. The output of the generator is the fake/estimated dielectric profile $G(\mathcal{N}+E_{z})=\widehat{\epsilon}_{r}(x,y)$. We refer to these dielectric profiles as fake since they are obtained as the output of a neural network instead of actual ground truth data. Since the exact thickness of the wall is not known, the dielectric profile is estimated over a specified Cartesian space that spans a greater spatial extent than the actual wall. The fake dielectric profile forms one of the two inputs to the discriminator. The other input to the discriminator are the corresponding actual/real dielectric profiles $\epsilon_r(x,y)$ of walls. Note that both the fake and real dielectric profiles are functions of the 2D Cartesian space. 
The objective of the generator is to map $\widehat{\epsilon}_r$ to $E_z$ while implicitly training the distribution of the fake dielectric profile samples to closely resemble the distribution of the real dielectric profile samples so that the generator can produce realistic estimates of the wall characteristics for a given test input electric field. The objective of the discriminator, on the other hand, is to distinguish between the real and fake samples.
Together, both networks of the GAN work in an adversarial manner where the weights of the generator and discriminator are optimized based on a value function $V(G_{\theta},D_{\phi})$ as given in:
\begin{multline}
\label{eqn:objectivefn}
\min_{G_{\theta}} \max_{D_{\phi}} V(G_{\theta},D_{\phi})=\min_{G_{\theta}} \max_{D_{\phi}} log(D_{\phi}(\epsilon_{r})) \\ +\log(1-D_{\phi}(G_{\theta}(\mathcal{N}+E_{z})))
\end{multline}
While training the generator, the weights and biases of the generator network are updated by keeping the discriminator constant. Similarly, while training the discriminator network, the weights and bias of the discriminator network are updated, keeping the generator weights constant. The training process involves iterative simultaneous stochastic gradient descent based on Adam optimization \cite{kingma2014adam}. Since the loss function of each network depends on the other network’s parameters, but each network cannot control the other network’s parameters, the scenario is a zero-sum game. During the test stage, we provide test electric field $\widehat E_{z}$ along with random noise to the generator to estimate the actual dielectric profile $\widehat{\epsilon}_{r}$ as shown in Fig.\ref{fig:Training}b. This is compared with ground truth data for validation.
\subsection{Network details}
We implement the GAN-based electromagnetic inversion through multiple approaches and evaluate the effectiveness of each approach.\\
\textbf{GAN-ANNf:} In the first method, we configure the two networks with artificial neural networks (ANN). We consider narrowband frequency domain complex-valued scattered electric field at the receivers as input to the GAN. The generator network is configured with two hidden layers with 256 and 512 hidden nodes and the \emph{leaky ReLU} as activation function. The $tanh$ activation function is used in the output layer with 1024 nodes. The discriminator network consists of two hidden layers with 512 and 256 hidden nodes, with \emph{leaky ReLU} activation functions. The output layer has one node with a $sigmoid$ function. \\
\textbf{GAN-ANNt:} In the second approach, we use time-domain electric field data as input to the generator. The generator consists of two hidden layers with 512 and 768 hidden nodes, while the discriminator architecture is same as GAN-ANNf. \\
\textbf{GAN-CNNf:} In the next approach, the ANN in the generator and discriminator are replaced by convolutional neural networks (CNN) while the frequency domain electric field data are retained as input to the generator. Here, we have two hidden layers with 128 filters each of $4 \times 4$ kernel size with a stride of $2\times 2$ in the generator, while in the discriminator, we have two hidden layers with 128 filters each of $3 \times 3$ kernel size with a stride of $2\times 2$. \\
\textbf{GAN-CNNt:} In the fourth and final approach, we configure the two GAN networks with CNN and time-domain electric field as input to the generator. In the generator, the \emph{Leaky ReLU} function is used for the hidden layers and $tanh$ for the output layer. The generator has two hidden layers with 128 filters each. The first layer consists of $5 \times 5$ kernel size with a stride of $1\times 1$ while the second layer consists of $3 \times 3$  kernel size with a stride of $2\times 2$. The discriminator architecture is same as GAN-CNNf. \\
The networks are configured with 0.0002 learning rate and 0.2 dropout rate for all the cases. The network parameters discussed here are chosen based on thumb rules with respect to the size of the input and output vectors to the GAN and experimental results.  The full details of the experiments are presented in the appendix.
\color{black}
\section{Simulation Setup}
\label{sec:Sim}
\begin{figure*}[htbp]
\centering
\includegraphics[scale=0.2]{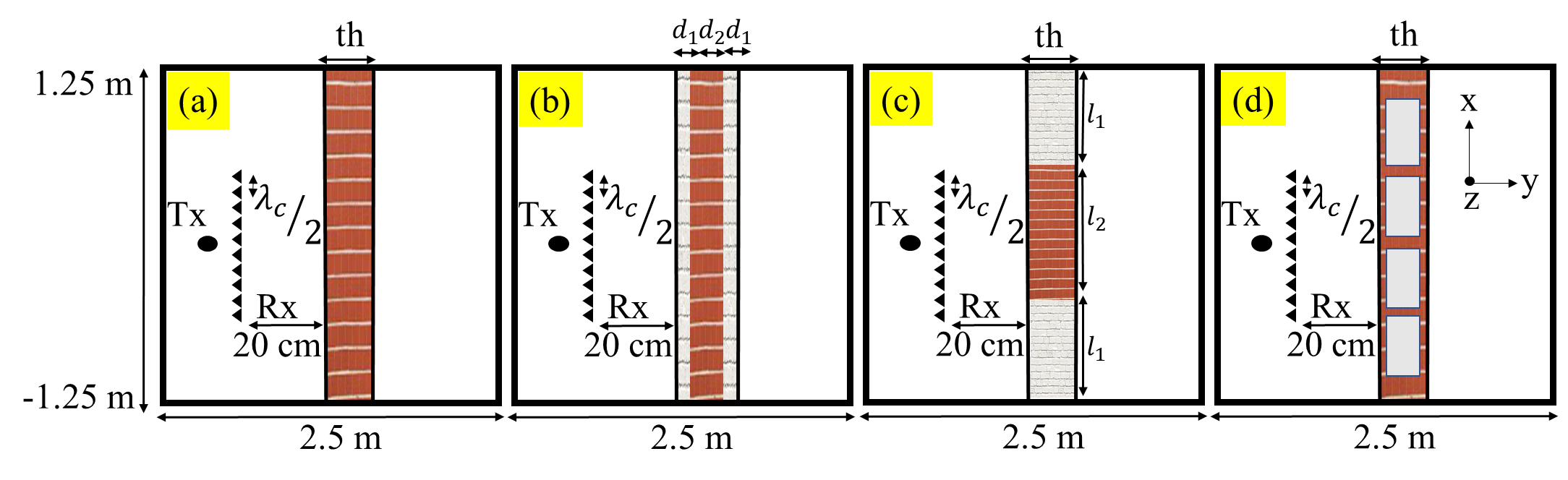}
\caption{FDTD simulation setup for single transmitter and multiple element receiver arrays before (a) a homogeneous dielectric wall, (b) a $y$ layered wall, (c) an $x$ layered wall, and (d) dielectric wall with airgaps.}
\label{fig:simulation_setup}
\end{figure*}

In this section, we describe the simulation method to generate scattered electric field data from different types of walls. The  scattering phenomenology is modeled using finite-difference time-domain (FDTD) techniques \cite{yee1966numerical}. We consider a 2D simulation space spanning -1.25m to 1.25m along $x$ and -0.25m to 2.25m along $y$ directions as shown in Fig.\ref{fig:simulation_setup}. 
We consider four types of walls that are assumed to be uniform along $z$. All of these walls resemble actual walls encountered in real-world scenarios. For each of these wall types, we consider multiple instances with distinct electrical characteristics as specified in Table.\ref{table:total_cases}.
The first wall type is a homogeneous dielectric wall whose relative permittivity ($\epsilon_r$) is varied uniformly from 3 to 8 to model the properties of common interior wall materials such as wood, brick, cement, and mud \cite{balanis2012advanced} while its thickness ($th$) is varied uniformly from 10cm to 50cm. The second wall type is an inhomogeneous wall with three dielectric layers aligned in the $y$ direction corresponding to a dielectric wall with insulation materials/facades on either side. The inner layer of thickness $d_2$ corresponds to a higher dielectric constant ($\epsilon_{r_2}$) than that of the two outer layers ($\epsilon_{r_1}$) of thickness $d_1$. The third type of wall are inhomogeneous walls with three dielectric layers aligned along the $x$ direction. This corresponds to dielectric walls with window or doorways. Again, the two outer layers are assumed to have identical length ($l_1$) and dielectric constant ($\epsilon_{r_1}$) while the inner layer of length $l_2$ has a higher dielectric constant ($\epsilon_{r_2}$). The thickness of the wall ($th$ along $y$) is also varied in this case. The last wall type is a dielectric wall with periodic air gaps corresponding to cinderblock walls. In this case, the dielectric constant of the wall material $\epsilon_r$ is varied uniformly from 3 to 8 and the thickness, $th$, is varied from 20cm to 50cm. Further, we vary the total number of air gaps across the length of the wall from 2 to 4 while maintaining a constant edge thickness of 10cm.
\begin{table}[!htbp]
\centering
\caption{Dielectric constant and thicknesses of walls}
\label{table:total_cases}
\begin{tabular}{*5c}
\hline 
\noalign{\vskip 1pt}
    {Wall type}&{Parameter}&{Values}&{cases}&{total cases}\\
\hline 
\noalign{\vskip 1pt}
     {Homogeneous}&{$\epsilon_{r}$}&{3-8}&{26}&{}\\
     {}&{$th$}&   {10-50cm}& { 5}& {130}\\\hline  
{Y-layered wall}&{$\epsilon_{r_2}$}&  {4-8}&{5}&{}\\
  {}&            {$\epsilon_{r_1}$}&  {2-3}& {3}& {}\\
  {}&            {$d_2$}&  {10-30cm}& {5}& {} \\
  {}&            {$d_1$}&  {5-15cm}& {3}&{225}\\\hline 
 { X-layered wall}&{$\epsilon_{r_2}$}&  {4-8}& {5}& {}\\
  {}&            {$\epsilon_{r_1}$}&  {2-3}&{3}& {}\\
  {}&            {$l_2$}&  {60-80cm}& {5}& {}\\
  {}&            {$l_1$}&  {$(2-l_2)/2$}&{-}&{}\\
  {}&            {$th$}&  {10-50cm}& {5}& {225}\\\hline
  
 {Wall with air gaps}&  {$\epsilon_{r}$}& {3-8}&  {26}&{}\\
  {}&                 {$th$}&  {20-50cm}& {4}&{}\\
  {}&               {air gaps}&{2-4}&{3}&{312}\\
                       
     \hline 
\end{tabular}
\end{table}
Altogether, we consider 892 distinct wall cases and the details of the entire dataset are available at \url{https://essrg.iiitd.edu.in/?page_id=4355}.

We consider an infinitely long line source excitation corresponding to the transmitter at (0m,0.5m). The source consists of a sinusoidal excitation at 2.4GHz and gives rise to transverse magnetic wave propagation in the 2D space. The simulation space is bounded by a perfectly matched layer of $2\lambda_c$ thickness where $\lambda_c$ corresponds to the wavelength. The entire 2D space is discretized into uniform grids that are one-tenth the wavelength resolution. The electric field propagates from the source and impinges upon the wall. Some of the energy propagates through the wall while the remaining is scattered. The time-domain electric field is recorded along with a uniform linear array, at a specified standoff distance behind the front face of the wall, as indicated in the figure, where the elements are spaced half wavelength apart from -0.28m to +0.28m. The duration of the simulation is set as 21.5ns with a time resolution of 0.02ns to ensure Courant stability conditions \cite{zheng1999finite}. The FDTD operation is then repeated for free-space conditions with the same source excitation, and the electric field is recorded at the ten receiver positions. Then the electric field of the free space scenarios is subtracted from that of the wall scenarios in order to remove the direct coupling between the transmitter and the receivers.

The time-domain electric field data at each of the receiver positions are down-sampled by a factor of 20 to obtain 52-time samples. The down-sampling operation is carried out to reduce the complexity of the deep learning architecture. Note that since the source excitation is 2.4GHz, the down-sampling operation does not result in aliasing. The data from the 10 receiver positions are concatenated to form a single column vector of $[520 \times 1]$ size. This is concatenated with a latent Gaussian noise vector of $[2600 \times 1]$ size of zero mean and unity variance and provided as the input to the generator for the GAN-ANNt and GAN-CNNt. For the frequency domain variants of the GAN (GAN-ANNf and GAN-CNNf), we perform Fourier transform on the time-domain (before down-sampling) electric field at the ten receiver positions and then extract the complex components corresponding to 2.4GHz, which are provided as input to the generator. The real part of the fields is concatenated with the imaginary parts of the field to form a $[20 \times 1]$ vector, which is concatenated with a noise vector of size $[100 \times 1]$ and provided as input to the generator.

The dielectric profile corresponds to a spatial extent of the 2D Cartesian space spanning along $x$ from -1m to +1m and along $y$ from 1m to 1.8m with a spatial resolution of 0.01m. Note that the spatial resolution for the GAN is higher than the spatial resolution of the FDTD grid space and corresponds to $32 \times 32$ pixel size of the dielectric profile. The higher resolution profile of the FDTD space results in a much higher computational cost of the GAN and hence is not adopted. In ANN architectures (GAN-ANNt and GAN-ANNf) the output from the generator is mapped to the fake dielectric profile of size $[1024 \times 1]$ (2D matrix reshaped to a column), while in the CNN architectures (GAN-CNNt and GAN-CNNf), the output retains the 2D pixel structure. The spatial extent of the mapped dielectric profile in the $y$ direction of 0.8m is greater than the thickness of most real-world walls and chosen so as to enable the estimation of the actual wall thickness. Of the total 892 data samples, we divide the data set such that $90\%$ are used for training and the remaining $10\%$ for validation. The batch size is taken as 16, and the model is trained for 1000 epochs.
\section{Simulation Results}
In this section, we evaluate the effectiveness of the proposed algorithms for estimating the wall characteristics. We validate our GAN networks with test data - simulated electric field data ($\tilde{E_z}$) from walls with known dielectric profiles ($\epsilon_r$). The test data are fed to the generator from which we estimate the dielectric profile $\tilde{\epsilon_r}$ which is subsequently compared with the ground truth dielectric profile. 
\subsection{Comparison of GAN with DNNs and classical methods}
We evaluate the four approaches of GAN that we discussed in the previous section- GAN-ANNf, GAN-ANNt, GAN-CNNt and GAN-CNNf. We compare their performances with two other popular deep learning networks - the fully connected neural network (FC-NN) and the convolutional neural network (CNN). Both these architectures consist of only a single neural network as opposed to the two adversarial networks used in GAN. We use both time and frequency domain input for the FC-NN. We also compare the performances with three classical EIS methods - BAM, BIM and BP - for frequency domain input data. We evaluate the results both visually and quantitatively.

\begin{figure}[htbp]
\centering
\includegraphics[scale=0.2]{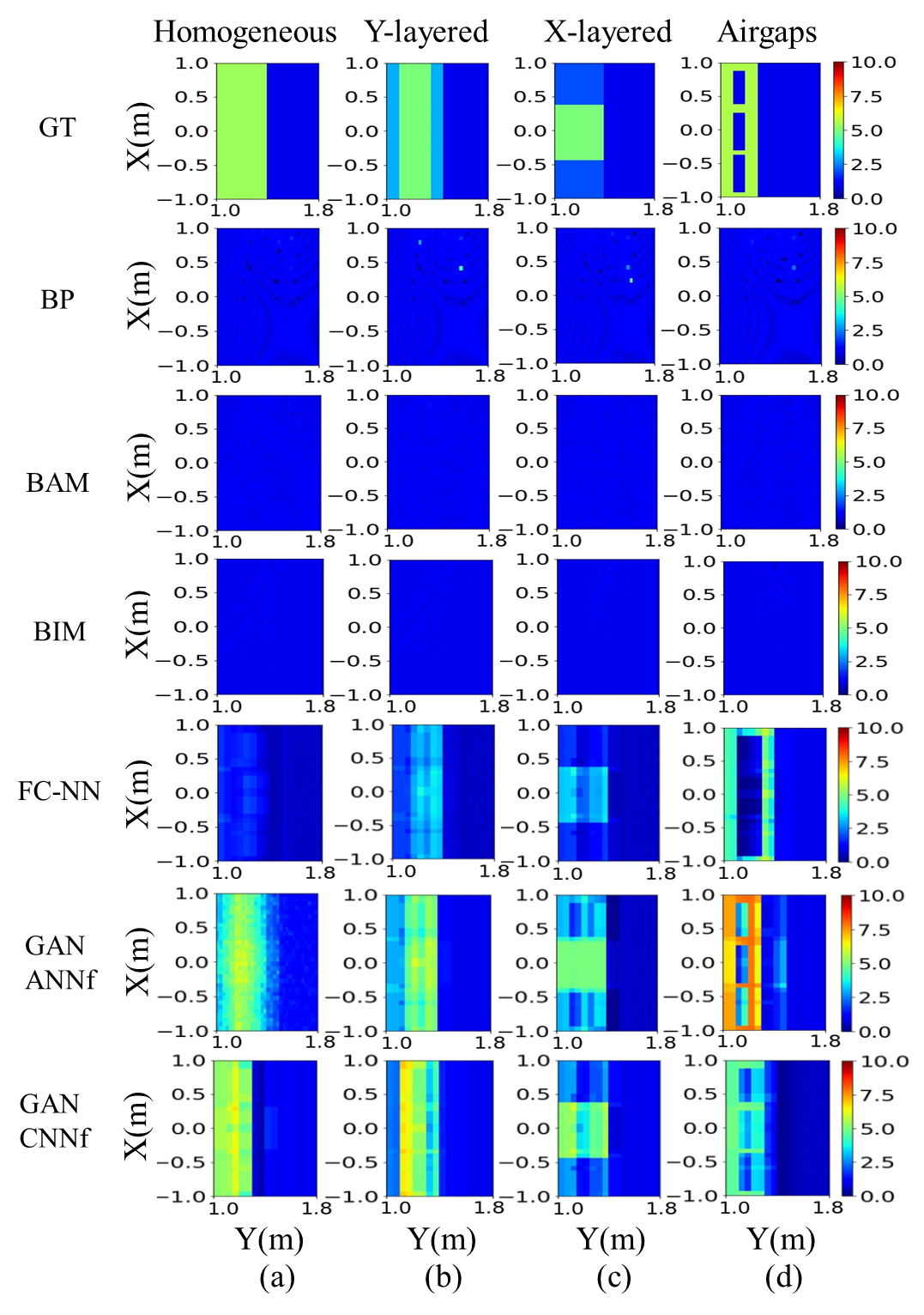}
\caption{Reconstructed dielectric profiles of homogeneous wall, $y$-layered wall, $x$-layered wall, and wall with air gaps from BP, BAM, BIM, FC-NN, GAN-ANNf and GAN-CNNf algorithms with frequency domain input data.}
\label{fig:Reconstruction_frequency_domain}
\end{figure}

\textbf{Frequency Domain:} First, we present the qualitative results for a test case for each wall type with frequency domain electric field data as input. In Fig.\ref{fig:Reconstruction_frequency_domain}, we present the dielectric profiles estimated for four wall cases - homogeneous, $y$-layered wall, $x$-layered wall, and wall with air gaps. For each case, the ground truth (GT) dielectric profiles are presented in the first row. In the instance presented in Fig.\ref{fig:Reconstruction_frequency_domain}a, the dielectric constant of a 40cm thick homogeneous wall is 5.4. In Fig.\ref{fig:Reconstruction_frequency_domain}b, $(\epsilon_{r_1}=3,l_1=10cm )$ and $(\epsilon_{r_2}= 5,l_2=25cm)$. In Fig.\ref{fig:Reconstruction_frequency_domain}c, $(\epsilon_{r_1}=2,d_1=60cm)$ and $(\epsilon_{r_2}= 5, d_2=80cm)$. In Fig.\ref{fig:Reconstruction_frequency_domain}d, the wall has a dielectric constant of 5.6 and is 30cm thick. It has 3 airgaps with an edge thickness of 10cm.
The second to fourth rows present the results obtained from BP, BAM and BIM. For all cases of the wall, the estimated dielectric constant lie between 1 and 2.5, which is well below the ground truth values, and the estimated thickness is not accurate. The results are most erroneous for the wall with air gaps. These results are consistent with the findings in the literature that report the poor performance of classical inverse scattering approaches at estimating high dielectric constants even in homogeneous conditions \cite{chen2018computational}. In the FC-NN results in the fifth row of Fig.\ref{fig:Reconstruction_frequency_domain},
we observe that the reconstructed dielectric profiles for all four cases are closer to ground truth than the classical techniques. The thicknesses of the walls are estimated correctly except for walls with air gaps. However the estimate of the dielectric constant shows considerable error. We consider the reconstructed profiles generated by GAN-ANNf in the sixth row of Fig.\ref{fig:Reconstruction_frequency_domain}. 
Here we can see that the dielectric profile and thickness of the homogeneous wall lie between 4.5 and 5, and the estimated thickness is mostly accurate with slight distortion. Further, in the case of the $y$-layered wall, we see that the thicknesses of layers are estimated accurately. In the case of $x$-layered wall, the thicknesses of the layers and dielectric profile are estimated correctly. Same is mostly true for the wall with air gaps with the dielectric constant estimate ranging 5-7 and thickness of 30cm, respectively. Next, we consider the reconstructed profiles generated by GAN-CNNf in the seventh row of Fig.\ref{fig:Reconstruction_frequency_domain}. Here we can see that the dielectric profile and thickness of the homogeneous wall lie between 5 and 6, and the estimated thickness is mostly accurate. Further, in the case of the $y$-layered wall, we see that the thicknesses of layers are accurate. In the case of the $x$-layered wall, the dielectric profile is estimated correctly. Also, the wall with air gaps was estimated accurately, having a dielectric constant ranging 5-6 and a thickness of 30cm, respectively.

\begin{figure}[htbp]
\centering
\includegraphics[scale=0.23]{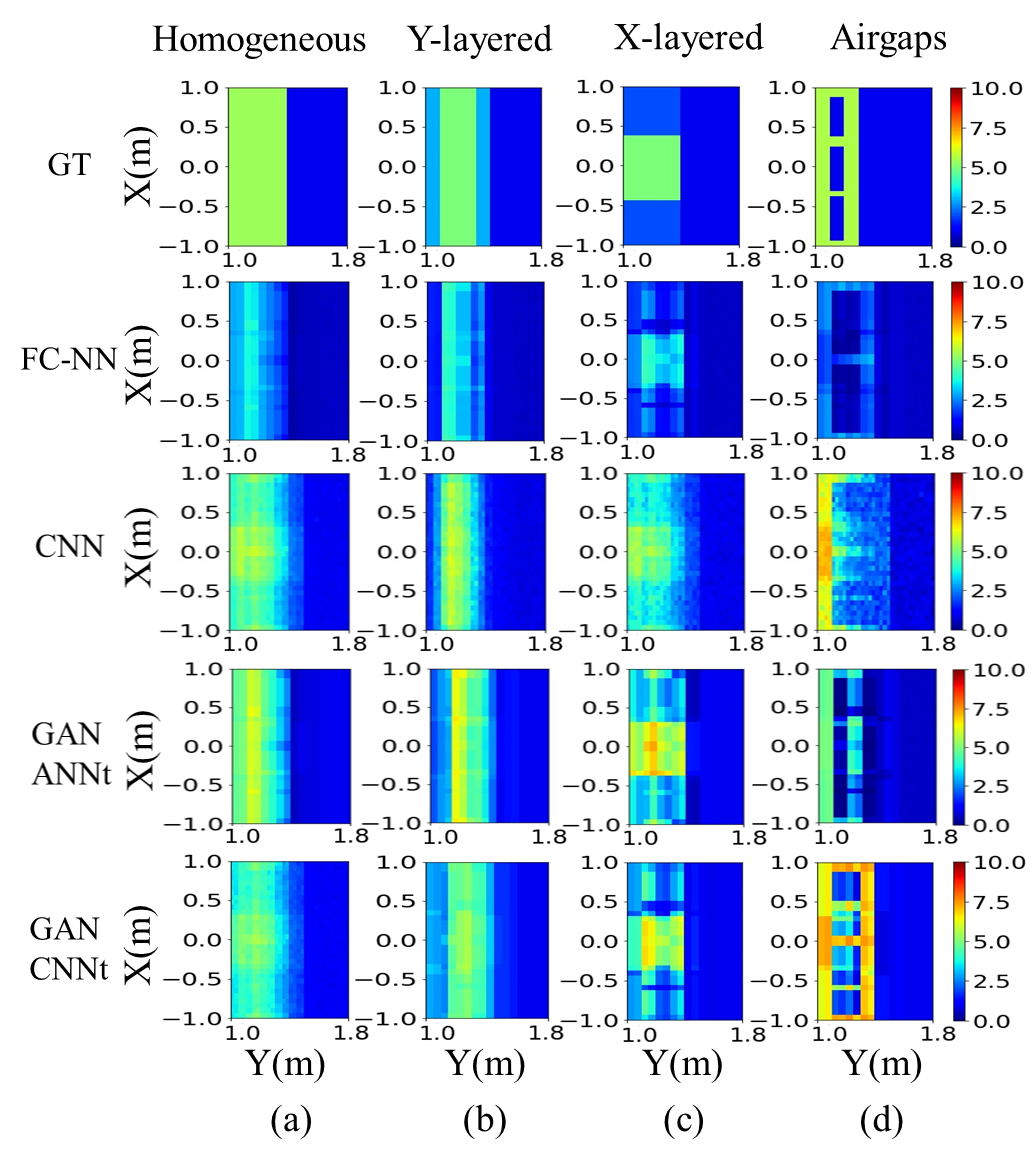}
\caption{Reconstructed dielectric profiles of homogeneous wall, $y$-layered wall, $x$-layered wall, and wall with air gaps from FC-NN,CNN,GAN-ANNt and GAN-CNNt, algorithms on time domain input data.}
\label{fig:Reconstruction_time_domain}
\end{figure}

\textbf{Time Domain:} Next, we consider the time domain data as input and consider the performance of the three algorithms FC-NN, GAN-ANNt and GAN-CNNt in Fig.\ref{fig:Reconstruction_time_domain}. 
The top row shows the ground truth for a single test case of each wall type. All three classical EM inversion techniques use the Green's function based in the frequency domain and hence cannot be used on the time-domain data. The results for FC-NN for all four wall cases show that the thickness of the wall is estimated somewhat correctly but the estimates of the dielectric constant are always well below the actual values. The CNN results in the third row show a slightly better performance but are still not able to estimate the thickness of the profiles accurately. For example, the air gaps in the last case are not accurately reconstructed. In the GAN-ANNt results in the fourth row of Fig.\ref{fig:Reconstruction_time_domain}, we see that the dielectric constant and thickness for the homogeneous wall are estimated correctly. In the case of the $x$-layered and $y$-layered walls, three distinct wall regions are correctly reconstructed and the thickness and dielectric constants are also fairly accurate. The estimated thickness of the wall with air gaps is lower than the ground truth. However, the airgaps are reconstructed correctly. We observe that the GAN-CNNt performs better in comparison to other methods in the fifth row of Fig.\ref{fig:Reconstruction_time_domain}.  Here we see that the dielectric constant is estimated accurately for all four cases. The same is also true for the thickness except for the wall with air gaps where the thickness estimate is a little higher than the ground truth. In conclusion, we observe visually that the GAN-based methods reconstruct the ground truth dielectric profiles more accurately than the classical methods and machine learning techniques with a single neural network.

For the quantitative comparison, the normalized mean square error (NMSE) is computed between the ground truth ($\epsilon_r$) and estimated dielectric profiles ($\tilde{\epsilon_r}$) using
\begin{equation}
    \label{eqn: NMSE}
   NMSE = \frac{\left \| \epsilon_r-\tilde{\epsilon_r}\right \|_{2}^{2}}{\left \|\epsilon_r  \right \|_{2}^{2}}.
\end{equation}
The NMSE is computed across 90 test cases of data comprising all four types of walls. The quantitative comparison of the techniques for all the test cases is summarized in Table.\ref{table:Comparison_table}.
\begin{table}[!htbp]
\centering
\caption{Benchmarking proposed GAN based inversion with classical EIS methods and single neural network methods.}
\label{table:Comparison_table}
\begin{tabular}{*5c}
\hline 
\noalign{\vskip 1pt}
    {Method}&{Input type (t/f)}&{NMSE}&{Training}&{Testing}\\
    {}&{}&{}&{Time(min)}&{Time(s/min)}\\
\hline 
\noalign{\vskip 1pt}
    {BP}&{f}&{0.74}&{-}&{20s}\\
  {BAM} &{f}&{0.83}&{-}&{3min}\\
     {BIM} &{f}&{0.8}&{-}&{91.5min}\\
     {FC-NN}&{t}&{0.42}& {30-45}&{0.1s}\\
     {FC-NN} &{f}&{0.4}&{30-45}&{0.1s}\\
     {CNN}&{t}&{0.25}&{30-45 }&{0.1s}\\
     {GAN-ANNf}&{f}&{0.23}&{120-150}&{0.015s}\\
     {GAN-ANNt}&{t}&{0.25}&{120-150}&{0.015s}\\
     {GAN-CNNf}&{f}&{0.2}&{150-180}&{0.015s}\\ 
     {GAN-CNNt}&{t}&{0.17}&{150-180}&{0.015s}\\
     \hline 
\end{tabular}
\end{table}
In Table.\ref{table:Comparison_table}, the input data to the algorithm is either time or frequency domain data denoted as \emph{t} or \emph{f} respectively.
The results show that the NMSE for the classical EIS techniques are much poorer than for the machine learning based techniques. The BP is slightly superior to the BAM. BIM introduces iterations in the dielectric profile estimation of the BAM algorithm. This improves the performance slightly. The classical  methods are unable to handle the high dielectric profiles in the scattering region. We also see that the proposed GAN methods are more accurate than FC-NN and CNN in obtaining lower NMSE, indicating that the use of two adversarial networks as opposed to one during training enables more accurate inversion operation. The GAN-CNNt network performs better than the same network with the frequency domain input (GAN-CNNf). It is possibly because of the information loss in the operation where the complex-valued data in the frequency domain are separated into real and imaginary parts and then concatenated and provided as input. Also, GAN-CNNt, GAN-CNNf and CNN networks achieve lower NMSE than their ANN counterparts indicating the strength of convolutional neural networks in capturing nonlinear relationships. 
\subsection{Computational Complexity}
We report the training and testing times for the algorithms in Table.\ref{table:Comparison_table}.
The computation cost in training and testing the learning algorithms arises from basic operations performed within the networks like convolution, addition, and calculation of activation functions. It also depends upon the number of layers, the types of layers, and the number of filters used in the neural networks. The classical EIS algorithms require no training. During test, the BIM is much longer than BAM due to the iterations within the algorithm till convergence is reached. In the case of the ANN architecture found in GAN-ANNt and GAN-ANNf, the complexity is of order $O(N^4)$ for $N$ stochastic gradient iterations for $N$ layers each with $N$ neurons and $N$ addition and multiplication operations \cite{zhou1988image}. The computation cost of FC-NN and CNN architecture is less than the GAN versions, which have two networks to train instead of one. Therefore, we note in Table.\ref{table:Comparison_table} the training times required by the GAN methods are comparatively greater than FC-NN and CNN. GAN-CNNt and GAN-CNNf has a CNN  architecture and its complexity is of $O(MNk^{2}IO)$ \cite{cong2014minimizing,wei2018deep}. Here, the number of the input and output feature maps are $I$ and $O$ respectively, $M \times N$ is the size of the feature map, and $k \times k$ is the convolution kernel size. Thus this algorithm takes the longest time to train. 
In the testing phase, the generator in the four different GAN methods is required to do a single forward process to obtain the reconstructed result without iterations. In our work, since the GAN architecture has few filters and hidden layers and because the multiple profiles are generated in parallel, the testing time for the GAN-based algorithm is lower than FC-NN and CNN. Thus the proposed techniques offer high accuracy with comparable low computational time during the test. The training and test codes are run with Keras 2.7 and trained and tested on an Intel Core i7-10510U processor running at 1.80 GHz. 
\subsection{Unexpected Test Conditions}
In all of our discussions so far, we have considered test scenarios that correspond to conditions that satisfy the initial assumptions in the problem statement.
Now, we study how the algorithm performs when unexpected scenarios are encountered during test conditions. Specifically, we consider two cases. First, when the wall is lossy and second is when a target is present along with the wall. 
\subsubsection{Lossy Walls:}
Most building wall materials are measured to have conductivities ranging from $0.01$ to $0.1S/m$ \cite{common2002propagation,stavrou2003review}.  For our test case, we consider the wall with air gaps with a thickness of 40cm having a dielectric constant varying uniformly between 4 and 8. We have taken five different values of conductivity, as shown in Table.\ref{table:Lossy_wall} and observed the outcome for the algorithms. 
\small
\begin{table}[!htbp]
\centering
\caption{Experimental results for lossy wall with airgaps}
\label{table:Lossy_wall}
{\begin{tabular}{*5c}
\hline 
\noalign{\vskip 1pt}
    Conductivity\\
    (S/m)&{GAN-CNNt}&{GAN-CNNf}&{GAN-ANNt}&{GAN-ANNf}\\
\hline 
\noalign{\vskip 1pt}
     {0}&{0.17}&{0.2}&{0.25}&{0.23}\\
     {0.001}&{0.53}&{0.48}&{0.57}&{0.56}\\
     {0.01}&{0.59}&{0.47}&{0.59}&{0.66}\\
     {0.1}&{0.67}&{0.54}&{0.83}&{0.79}\\
     {1}&{0.68}&{0.57}&{0.81}&{0.8}\\
     {10}&{0.94}&{0.93}&{1.2}&{0.9}\\  
     \hline 
\end{tabular}}
\end{table}
\normalsize
We observe that the errors are small when the conductivity is low - mirroring most real world conditions. However, 
as we increase the value of conductivity, the NMSE deteriorates for all four variants of the GAN algorithm. This loss in performance can be attributed to the narrowband scattered electric field data that are gathered in our method, which do not provide information regarding dispersive materials in the scattering region of interest. We anticipate an improvement in the performance when wideband data are used for inversion. 
\subsubsection{Presence of Target:}
\begin{figure}[htbp]
\centering
\includegraphics[scale= 0.45]{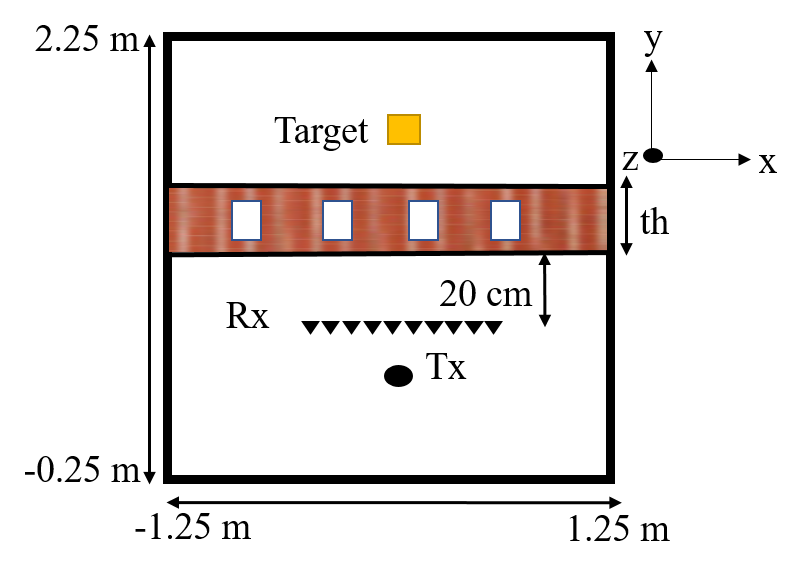}
\caption{Simulation setup with a dielectric target behind a wall with airgaps.}
\label{fig:wall_with_target}
\end{figure}

We have considered a rectangular target on the other side of the wall as shown in Fig.\ref{fig:wall_with_target}. 
We considered the wall with airgaps wall having dielectric value 4 and targets of five different sizes and at five different positions behind the wall resulting in a total of 25 test samples of scattered electric field from the wall with the presence of the target. We observe that the average NMSE for GAN-ANNf, GAN-ANNt, GAN-CNNt, and GAN-CNNf are 0.4, 0.42, 0.35 and 0.31 respectively. Hence, we can conclude that, in the presence of the target, our proposed algorithm's performance is slightly degraded but is still able to estimate the electrical characteristics of the walls. As previously noted, we observe that the GAN-CNNt and GAN-CNNf outperform the GAN-ANNf and GAN-ANNf due to the CNN architecture.
\color{black}
\section{Measurement Results}
\label{sec:Measurement}
In this section, we use the GAN trained with the simulated electric field data of the previous section to estimate the dielectric profiles of real walls. In other words, we use scattered electric field data measurements from actual walls as input to the generator, and the output is the estimate of the dielectric profile of the wall. Note that it is not possible to get the exact ground truth dielectric profile of the walls. Instead, we get a rough estimate based on the known materials used to construct the wall. 
\subsection{Measurement Setup}
\begin{figure}[htbp]
\centering
\includegraphics[width= 3.5in,height=2.5in]{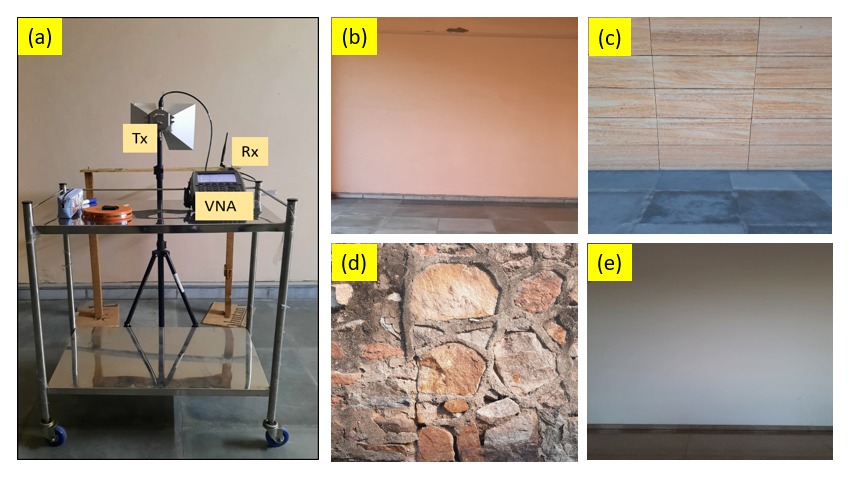}
\caption{(a) Experimental radar setup before different types of wall: (b) 30cm thick exterior brick and concrete wall, (c) 40cm thick exterior brick wall with a thin facade of ceramic tiles, (d) 40cm thick perimeter boundary wall made of large stones and concrete, and (e) 25cm thick interior brick and concrete wall.}
\label{fig:Experimental_setup}
\end{figure}

First, we describe the measurement setup used to collect narrowband scattered electric fields from different types of walls. The radar setup, shown in Fig.\ref{fig:Experimental_setup}a, consists of a two-port vector network analyzer Field Fox N9926A configured to make $S_{21}$ scattering parameter measurements at 2.4 GHz. 
The transmitted power is set at $3dBm$, and the sampling frequency of the measurement is 377Hz. We replicate the simulation set up as closely as possible by placing the transmitter 50 cm from the wall and gathering scattered electric fields from positions 20 cm before the front face of the wall. 
A linearly polarized broadband horn antenna (HF907) and a dipole antenna are connected to the two ports. The scattered signal from the wall is captured by the VNA, where it is amplified, inphase-quadrature demodulated and digitized. Then the complex VNA measurements are collected and downloaded to a laptop for further processing. We linearly move the dipole antenna to collect the data samples at ten positions corresponding to the receiver array discussed in the previous section. We consider four different walls, as shown in the figure. They are two exterior walls and one interior wall of a building, and one boundary wall from the perimeter of the campus, as shown in the figure. The first exterior wall is a 30cm thick brick wall with concrete and paint on either side. The second exterior wall is a 40cm brick wall with concrete on one side and a very thin layer of ceramic tiles having a thickness of 10mm forming a facade on the other side. The third wall is a perimeter boundary wall made of large stones and concrete. This is possibly the most inhomogeneous wall of the four, and its surface is rough, as shown in the figure. The last wall is a 25cm interior wall made of brick and concrete. The dielectric constant of brick lies between 3.8 and 4.2, stones lie between 6 and 8, concrete lies between 7.7 and 9.6, and ceramic tiles in between 21 and 28 respectively \cite{grosvenor2009time,stavrou2003review}. Measurements are repeated in outdoor free-space conditions. The narrowband $S_{21}$ measurements directly correspond to frequency-domain measurements. Two hundred samples are collected at each receiver position and averaged. Calibration of the wall scattered data is carried out by subtracting the free space results from the wall data to remove the effects of the direct coupling between the transmitter and the receiver. Since the measurements are made at each receiver position independently (synthetic array), there are no mutual coupling concerns in the data. 
\subsection{Results}
\begin{figure}[htbp]
\centering
\includegraphics[scale=0.21]{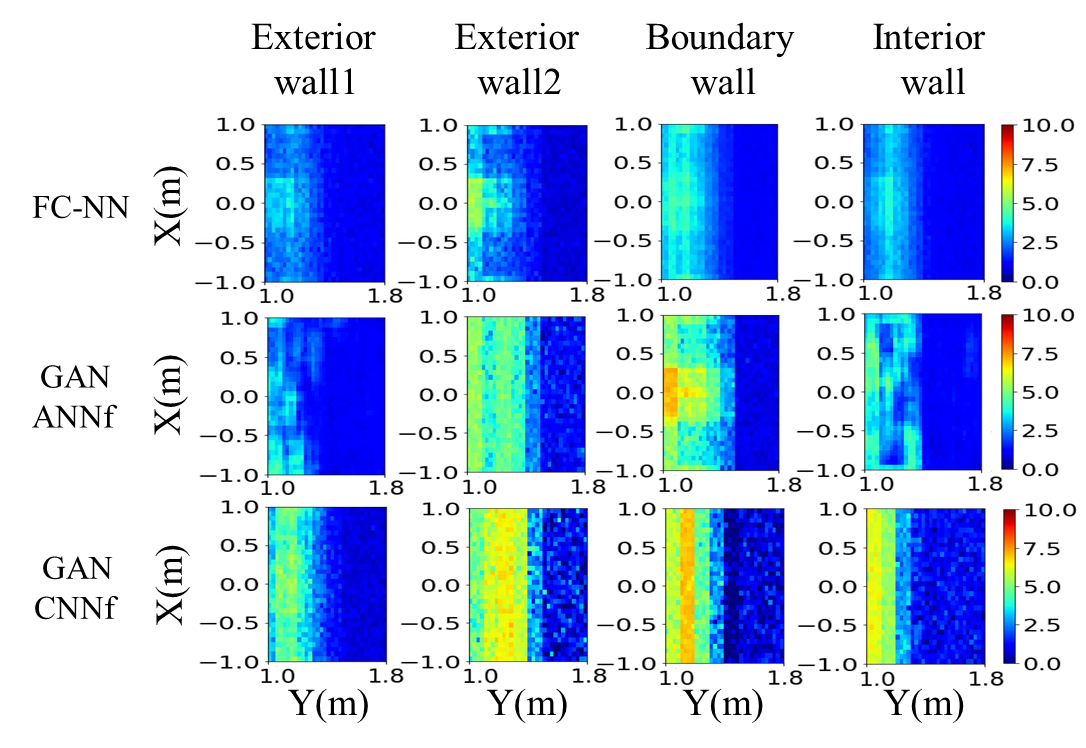}
\caption{Dielectric profiles of real walls reconstructed from neural networks: (a) 30cm thick exterior brick and concrete wall, (b) 40cm thick exterior brick wall with a thin facade of ceramic tiles, (c) 40cm thick perimeter boundary wall made of large stones and concrete, and (d) 25cm thick interior brick and concrete wall.}
\label{fig:Comparison_Experimental_wall_dielectric}
\end{figure}

Due to the nature of the VNA-based radar setup and its limited sampling frequency, it is not possible to carry out time-domain electric field measurements. Hence, we test the performance of the GAN-CNNf, GAN-ANNf and compare it with FC-NN. Due to the poor performance of the classical EIS methods on ideal simulated data, we do not use implement those methods with the measured data gathered from walls with high dielectric profiles. The reconstructed dielectric profiles of the four walls are shown in Fig.\ref{fig:Comparison_Experimental_wall_dielectric}.
The dielectric profile is reconstructed from the neural networks trained with simulation data. The first row shows the reconstructed dielectric profiles from FC-NN while the second row shows the results from GAN-ANNf and the third row shows the results from GAN-CNNf. The result shows that the GAN-based method is able to reconstruct the dielectric profile for the two exterior and one interior wall fairly well, unlike the FC-NN. The most challenging case is the perimeter stone wall which is highly inhomogeneous. The quantitative results are presented in Table.\ref{table:experimental_wall_characteristics}. 
\begin{table*}[!htbp]
\centering
\caption{The actual and estimated values of dielectric constant and thicknesses of real walls using FC-NN, GAN-ANNf and GAN-CNNf}
\label{table:experimental_wall_characteristics}
\begin{tabular}{c|ccc|cccc}
\hline
Wall type &  \multicolumn{3}{c}{Dielectric constant ($\epsilon_{r}$)} & \multicolumn{3}{c}{Thickness (cm)}\\

{} &FC-NN  &GAN-ANNf &GAN-CNNf &Actual &FC-NN  &GAN-ANNf&GAN-CNNf \\
\hline
Exterior Wall1\\
(Concrete, brick) &1-5.5&1-4&1-4                  &30 &10-30 &30-31 &20-30\\\hline
Exterior Wall2\\(Concrete, brick and tiles)
& 1-4.5    &1.3-6 &1.5-6                    &40 &5-20 &40-45&40-42\\\hline
Boundary Wall\\(stone, concrete) & 1.5-4.5   &3.5-8 &4-8              &40 &15-35 &41-45 &40-43\\\hline
Interior Wall\\(Concrete, brick) &  1.2-4.7   &3-6.2 &4-6               &25 &5-20  &25-30 &25-27\\
\hline
\end{tabular}
\end{table*}
Here, we observe that GAN-ANNf estimated the thickness of the walls with an error ranging from 2.5\% to 12.5\%, respectively, and the corresponding range of the dielectric constants. In comparison, the error in the estimates with FC-NN is very high (above 100\%) in some cases. This demonstrates that FC-NN is not able to handle scenarios where the test data (in this case, the measurement data) varies significantly from the training data (simulation data). 
\subsection{Discussion}
\begin{figure}[htbp]
\centering
\includegraphics[scale=0.3]{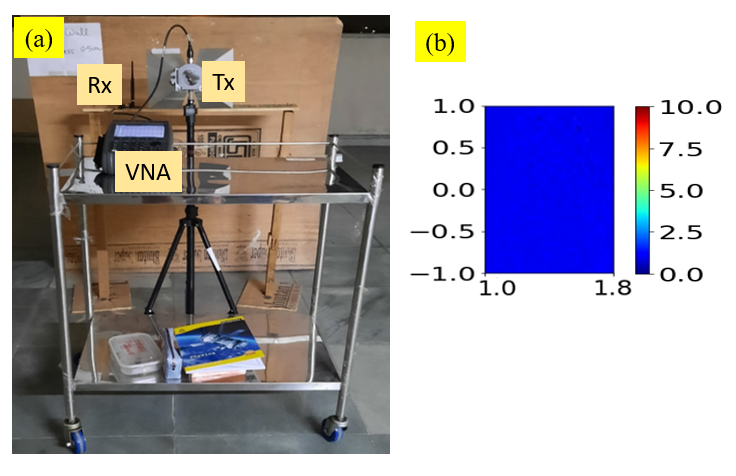}
\caption{(a) Experimental radar setup before 0.5cm thick plywood sheet. (b) Reconstructed dielectric profile of the plywood.}
\label{fig:Plywood_dielectric_profile}
\end{figure}

The performance of the method depends on the signal-to-noise ratio of the scattered electric field and the frequency of the source excitation. Some types of walls are poor reflectors at low microwave frequencies resulting in very weak scattered fields. To study this phenomenon, we repeated the exercise on an 0.5cm thick plywood wall as shown in Fig. \ref{fig:Plywood_dielectric_profile}a.
In this case, we are unable to reconstruct the dielectric profile as shown in Fig. \ref{fig:Plywood_dielectric_profile}b. This is because the plywood sheet is mostly transparent to the electromagnetic field at 2.4GHz, resulting in almost negligible scattered returns. A simple method to overcome this limitation is to increase the frequency of the source excitation. Fortunately, such walls do not cause significant distortions to through-wall radar signatures, and hence it is not imperative to remove the wall effects.

In all of our studies, the walls have been assumed to be uniform along with the height. Hence, the scattered electric field have been gathered across a linear array to capture the diversity in the wall characteristics across the two-dimensional Cartesian space. In scenarios where it is anticipated to have non-uniform characteristics across the height, the receivers can be placed along different heights to capture the effects of inhomogeneity along with the height. Further, we have assumed that the dielectric constant of the walls is real. When the dielectric constant is complex, the wall properties become dispersive. We believe the complex electrical characteristics can be estimated from scattered fields that arise from wideband source excitation instead of the narrowband source excitation assumed in this work. The study of more complex walls and the corresponding excitation will form the focus of future works. 
\section{Conclusions}
\label{sec:Conclusion}
In this paper, we proposed GAN-based approaches configured with time/frequency domain input scattered electric field data and ANN or CNN hidden layers to solve the EIS problem to estimate wall characteristics from scattered electric fields. We experimentally validate these approaches using both simulations and measurements. The results clearly demonstrate the effectiveness of the proposed methods in reconstructing high dielectric profiles in the scattering region of interest when compared to classical EIS methods. 
The use of two neural networks in an adversarial manner results in longer training times than other deep learning-based algorithms with single neural networks. However, they enable accurate and fast reconstruction of dielectric profiles of walls during the test. Further, the use of convolutional filters within the neural networks increases the training time but improves the accuracy during testing. The main advantage offered by the GAN-based algorithms is the limited requirement of a large volume of diverse training data. This means that the algorithm can be tested on real-world walls that are significantly different from those used during training. Our key contribution was to show how neural networks trained with simulated electric field data are useful for estimating the dielectric profiles and thickness of actual walls. However, when the walls are thin and made of materials that are transparent to the source excitation frequency, then the scattered fields are weak, and reconstruction is not possible. Fortunately, these types of walls do not significantly distort radar signatures and do not require reconstruction. 
\bibliographystyle{ieeetran}
\bibliography{main}

\begin{thebibliography}{10}
\providecommand{\url}[1]{#1}
\csname url@samestyle\endcsname
\providecommand{\newblock}{\relax}
\providecommand{\bibinfo}[2]{#2}
\providecommand{\BIBentrySTDinterwordspacing}{\spaceskip=0pt\relax}
\providecommand{\BIBentryALTinterwordstretchfactor}{4}
\providecommand{\BIBentryALTinterwordspacing}{\spaceskip=\fontdimen2\font plus
\BIBentryALTinterwordstretchfactor\fontdimen3\font minus
  \fontdimen4\font\relax}
\providecommand{\BIBforeignlanguage}[2]{{%
\expandafter\ifx\csname l@#1\endcsname\relax
\typeout{** WARNING: IEEEtran.bst: No hyphenation pattern has been}%
\typeout{** loaded for the language `#1'. Using the pattern for}%
\typeout{** the default language instead.}%
\else
\language=\csname l@#1\endcsname
\fi
#2}}
\providecommand{\BIBdecl}{\relax}
\BIBdecl

\bibitem{amin2014through}
M.~G. Amin and F.~Ahmad, ``Through-the-wall radar imaging: theory and
  applications,'' in \emph{Academic Press Library in Signal Processing}.\hskip
  1em plus 0.5em minus 0.4em\relax Elsevier, 2014, vol.~2, pp. 857--909.

\bibitem{amin2016radar}
M.~G. Amin, Y.~D. Zhang, F.~Ahmad, and K.~D. Ho, ``Radar signal processing for
  elderly fall detection: The future for in-home monitoring,'' \emph{IEEE
  Signal Processing Magazine}, vol.~33, no.~2, pp. 71--80, 2016.

\bibitem{ram2008doppler}
S.~S. Ram, Y.~Li, A.~Lin, and H.~Ling, ``Doppler-based detection and tracking
  of humans in indoor environments,'' \emph{Journal of the Franklin Institute},
  vol. 345, no.~6, pp. 679--699, 2008.

\bibitem{ram2010simulation}
S.~S. Ram, C.~Christianson, Y.~Kim, and H.~Ling, ``Simulation and analysis of
  human micro-dopplers in through-wall environments,'' \emph{IEEE Transactions
  on Geoscience and remote sensing}, vol.~48, no.~4, pp. 2015--2023, 2010.

\bibitem{ram2009simulation}
S.~S. Ram, C.~Christianson, and H.~Ling, ``Simulation of high range-resolution
  profiles of humans behind walls,'' in \emph{2009 IEEE Radar
  Conference}.\hskip 1em plus 0.5em minus 0.4em\relax IEEE, 2009, pp. 1--4.

\bibitem{qi2016detection}
F.~Qi, F.~Liang, H.~Lv, C.~Li, F.~Chen, and J.~Wang, ``Detection and
  classification of finer-grained human activities based on stepped-frequency
  continuous-wave through-wall radar,'' \emph{Sensors}, vol.~16, no.~6, p. 885,
  2016.

\bibitem{narayanan2008through}
R.~M. Narayanan, ``Through-wall radar imaging using uwb noise waveforms,''
  \emph{Journal of the Franklin Institute}, vol. 345, no.~6, pp. 659--678,
  2008.

\bibitem{lai2010ultrawideband}
C.-P. Lai and R.~M. Narayanan, ``Ultrawideband random noise radar design for
  through-wall surveillance,'' \emph{IEEE Transactions on Aerospace and
  Electronic Systems}, vol.~46, no.~4, pp. 1716--1730, 2010.

\bibitem{ram2007human}
S.~S. Ram, Y.~Li, A.~Lin, and H.~Ling, ``Human tracking using doppler
  processing and spatial beamforming,'' in \emph{2007 IEEE Radar
  Conference}.\hskip 1em plus 0.5em minus 0.4em\relax IEEE, 2007, pp. 546--551.

\bibitem{ram2008through}
S.~S. Ram and H.~Ling, ``Through-wall tracking of human movers using joint
  doppler and array processing,'' \emph{IEEE Geoscience and Remote Sensing
  Letters}, vol.~5, no.~3, pp. 537--541, 2008.

\bibitem{ahmad2005synthetic}
F.~Ahmad, M.~G. Amin, and S.~A. Kassam, ``Synthetic aperture beamformer for
  imaging through a dielectric wall,'' \emph{IEEE transactions on aerospace and
  electronic systems}, vol.~41, no.~1, pp. 271--283, 2005.

\bibitem{ahmad2008three}
F.~Ahmad, Y.~Zhang, and M.~G. Amin, ``Three-dimensional wideband beamforming
  for imaging through a single wall,'' \emph{IEEE Geoscience and remote sensing
  letters}, vol.~5, no.~2, pp. 176--179, 2008.

\bibitem{solimene2009three}
R.~Solimene, F.~Soldovieri, G.~Prisco, and R.~Pierri, ``Three-dimensional
  through-wall imaging under ambiguous wall parameters,'' \emph{IEEE
  Transactions on Geoscience and Remote Sensing}, vol.~47, no.~5, pp.
  1310--1317, 2009.

\bibitem{solimene2011twi}
R.~Solimene, R.~Di~Napoli, F.~Soldovieri, and R.~Pierri, ``Twi for an unknown
  symmetric lossless wall,'' \emph{IEEE Transactions on Geoscience and Remote
  Sensing}, vol.~49, no.~8, pp. 2876--2886, 2011.

\bibitem{wang2017look}
X.~Wang, G.~Li, Q.~Wan, and R.~J. Burkholder, ``Look-ahead hybrid matching
  pursuit for multipolarization through-wall radar imaging,'' \emph{IEEE
  Transactions on Geoscience and Remote Sensing}, vol.~55, no.~7, pp.
  4072--4081, 2017.

\bibitem{guo2017imaging}
S.~Guo, G.~Cui, L.~Kong, and X.~Yang, ``An imaging dictionary based multipath
  suppression algorithm for through-wall radar imaging,'' \emph{IEEE
  Transactions on Aerospace and Electronic Systems}, vol.~54, no.~1, pp.
  269--283, 2017.

\bibitem{liu2019clutter}
H.~Liu, C.~Huang, L.~Gan, Y.~Zhou, and T.-K. Truong, ``Clutter reduction and
  target tracking in through-the-wall radar,'' \emph{IEEE Transactions on
  Geoscience and Remote Sensing}, vol.~58, no.~1, pp. 486--499, 2019.

\bibitem{vishwakarma2020mitigation}
S.~Vishwakarma and S.~S. Ram, ``Mitigation of through-wall distortions of
  frontal radar images using denoising autoencoders,'' \emph{IEEE Transactions
  on Geoscience and Remote Sensing}, vol.~58, no.~9, pp. 6650--6663, 2020.

\bibitem{ram2021sparsity}
S.~S. Ram, S.~Vishwakarma, A.~Sneh, and K.~Yasmeen, ``Sparsity based
  autoencoders for denoising cluttered radar signatures,'' \emph{IET Radar,
  Sonar \& Navigation}, vol.~15, no.~8, pp. 915--931, 2021.

\bibitem{setlur2011multipath}
P.~Setlur, M.~Amin, and F.~Ahmad, ``Multipath model and exploitation in
  through-the-wall and urban radar sensing,'' \emph{IEEE Transactions on
  Geoscience and Remote Sensing}, vol.~49, no.~10, pp. 4021--4034, 2011.

\bibitem{ghorbani2021deep}
F.~Ghorbani, H.~Soleimani, and M.~Soleimani, ``Deep learning approach for
  target locating in through-the-wall radar under electromagnetic complex
  wall,'' \emph{arXiv preprint arXiv:2102.07990}, 2021.

\bibitem{guo2018multipath}
S.~Guo, X.~Yang, G.~Cui, Y.~Song, and L.~Kong, ``Multipath ghost suppression
  for through-the-wall imaging radar via array rotating,'' \emph{IEEE
  Geoscience and Remote Sensing Letters}, vol.~15, no.~6, pp. 868--872, 2018.

\bibitem{zheng2018multilevel}
C.~Zheng, X.~Xi, Z.~Song, and K.~Zhang, ``Multilevel delay lock loop approach
  for wall clutter mitigation in through-the-wall radar imaging,'' \emph{IET
  Microwaves, Antennas \& Propagation}, vol.~12, no.~12, pp. 1986--1992, 2018.

\bibitem{soldovieri2007through}
F.~Soldovieri and R.~Solimene, ``Through-wall imaging via a linear inverse
  scattering algorithm,'' \emph{IEEE Geoscience and Remote Sensing Letters},
  vol.~4, no.~4, pp. 513--517, 2007.

\bibitem{habashy1993beyond}
T.~M. Habashy, R.~W. Groom, and B.~R. Spies, ``Beyond the born and rytov
  approximations: A nonlinear approach to electromagnetic scattering,''
  \emph{Journal of Geophysical Research: Solid Earth}, vol.~98, no.~B2, pp.
  1759--1775, 1993.

\bibitem{chen2018computational}
X.~Chen, \emph{Computational methods for electromagnetic inverse
  scattering}.\hskip 1em plus 0.5em minus 0.4em\relax John Wiley \& Sons, 2018.

\bibitem{belkebir2005superresolution}
K.~Belkebir, P.~C. Chaumet, and A.~Sentenac, ``Superresolution in total
  internal reflection tomography,'' \emph{JOSA A}, vol.~22, no.~9, pp.
  1889--1897, 2005.

\bibitem{wang1989iterative}
Y.~Wang and W.~C. Chew, ``An iterative solution of the two-dimensional
  electromagnetic inverse scattering problem,'' \emph{International Journal of
  Imaging Systems and Technology}, vol.~1, no.~1, pp. 100--108, 1989.

\bibitem{chew1990reconstruction}
W.~C. Chew and Y.-M. Wang, ``Reconstruction of two-dimensional permittivity
  distribution using the distorted born iterative method,'' \emph{IEEE
  transactions on medical imaging}, vol.~9, no.~2, pp. 218--225, 1990.

\bibitem{song2005through}
L.-P. Song, C.~Yu, and Q.~H. Liu, ``Through-wall imaging (twi) by radar: 2-d
  tomographic results and analyses,'' \emph{IEEE Transactions on Geoscience and
  Remote Sensing}, vol.~43, no.~12, pp. 2793--2798, 2005.

\bibitem{chen2009subspace}
X.~Chen, ``Subspace-based optimization method for solving inverse-scattering
  problems,'' \emph{IEEE Transactions on Geoscience and Remote Sensing},
  vol.~48, no.~1, pp. 42--49, 2009.

\bibitem{zhong2011fft}
Y.~Zhong and X.~Chen, ``An fft twofold subspace-based optimization method for
  solving electromagnetic inverse scattering problems,'' \emph{IEEE
  transactions on antennas and propagation}, vol.~59, no.~3, pp. 914--927,
  2011.

\bibitem{semnani2009reconstruction}
A.~Semnani, M.~Kamyab, and I.~T. Rekanos, ``Reconstruction of one-dimensional
  dielectric scatterers using differential evolution and particle swarm
  optimization,'' \emph{IEEE Geoscience and Remote Sensing Letters}, vol.~6,
  no.~4, pp. 671--675, 2009.

\bibitem{chen2020review}
X.~Chen, Z.~Wei, M.~Li, and P.~Rocca, ``A review of deep learning approaches
  for inverse scattering problems (invited review),'' \emph{Progress In
  Electromagnetics Research}, vol. 167, pp. 67--81, 2020.

\bibitem{caorsi1999electromagnetic}
S.~Caorsi and P.~Gamba, ``Electromagnetic detection of dielectric cylinders by
  a neural network approach,'' \emph{IEEE transactions on geoscience and remote
  sensing}, vol.~37, no.~2, pp. 820--827, 1999.

\bibitem{rekanos2002neural}
I.~T. Rekanos, ``Neural-network-based inverse-scattering technique for online
  microwave medical imaging,'' \emph{IEEE transactions on magnetics}, vol.~38,
  no.~2, pp. 1061--1064, 2002.

\bibitem{cicchetti2021numerical}
R.~Cicchetti, S.~Pisa, E.~Piuzzi, E.~Pittella, P.~D'Atanasio, and O.~Testa,
  ``Numerical and experimental comparison among a new hybrid ft-music technique
  and existing algorithms for through-the-wall radar imaging,'' \emph{IEEE
  Transactions on Microwave Theory and Techniques}, 2021.

\bibitem{zheng2021human}
Z.~Zheng, J.~Pan, Z.~Ni, C.~Shi, S.~Ye, and G.~Fang, ``Human posture
  reconstruction for through-the-wall radar imaging using convolutional neural
  networks,'' \emph{IEEE Geoscience and Remote Sensing Letters}, 2021.

\bibitem{li2020human}
H.~Li, G.~Cui, S.~Guo, L.~Kong, and X.~Yang, ``Human target detection based on
  fcn for through-the-wall radar imaging,'' \emph{IEEE Geoscience and Remote
  Sensing Letters}, 2020.

\bibitem{goodfellow2020generative}
I.~Goodfellow, J.~Pouget-Abadie, M.~Mirza, B.~Xu, D.~Warde-Farley, S.~Ozair,
  A.~Courville, and Y.~Bengio, ``Generative adversarial networks,''
  \emph{Communications of the ACM}, vol.~63, no.~11, pp. 139--144, 2020.

\bibitem{radford2015unsupervised}
A.~Radford, L.~Metz, and S.~Chintala, ``Unsupervised representation learning
  with deep convolutional generative adversarial networks,'' \emph{arXiv
  preprint arXiv:1511.06434}, 2015.

\bibitem{salimans2016improved}
T.~Salimans, I.~Goodfellow, W.~Zaremba, V.~Cheung, A.~Radford, and X.~Chen,
  ``Improved techniques for training gans,'' in \emph{Advances in neural
  information processing systems}, 2016, pp. 2234--2242.

\bibitem{creswell2018generative}
A.~Creswell, T.~White, V.~Dumoulin, K.~Arulkumaran, B.~Sengupta, and A.~A.
  Bharath, ``Generative adversarial networks: An overview,'' \emph{IEEE Signal
  Processing Magazine}, vol.~35, no.~1, pp. 53--65, 2018.

\bibitem{hinton2006fast}
G.~E. Hinton, S.~Osindero, and Y.-W. Teh, ``A fast learning algorithm for deep
  belief nets,'' \emph{Neural computation}, vol.~18, no.~7, pp. 1527--1554,
  2006.

\bibitem{liu2018generative}
Z.~Liu, D.~Zhu, S.~P. Rodrigues, K.-T. Lee, and W.~Cai, ``Generative model for
  the inverse design of metasurfaces,'' \emph{Nano letters}, vol.~18, no.~10,
  pp. 6570--6576, 2018.

\bibitem{kingma2014adam}
D.~P. Kingma and J.~Ba, ``Adam: A method for stochastic optimization,''
  \emph{arXiv preprint arXiv:1412.6980}, 2014.

\bibitem{yee1966numerical}
K.~Yee, ``Numerical solution of initial boundary value problems involving
  maxwell's equations in isotropic media,'' \emph{IEEE Transactions on antennas
  and propagation}, vol.~14, no.~3, pp. 302--307, 1966.

\bibitem{balanis2012advanced}
C.~A. Balanis, \emph{Advanced engineering electromagnetics}.\hskip 1em plus
  0.5em minus 0.4em\relax John Wiley \& Sons, 2012.

\bibitem{zheng1999finite}
F.~Zheng, Z.~Chen, and J.~Zhang, ``A finite-difference time-domain method
  without the courant stability conditions,'' \emph{IEEE Microwave and Guided
  wave letters}, vol.~9, no.~11, pp. 441--443, 1999.

\bibitem{zhou1988image}
Y.-T. Zhou, R.~Chellappa, A.~Vaid, and B.~K. Jenkins, ``Image restoration using
  a neural network,'' \emph{IEEE transactions on acoustics, speech, and signal
  processing}, vol.~36, no.~7, pp. 1141--1151, 1988.

\bibitem{cong2014minimizing}
J.~Cong and B.~Xiao, ``Minimizing computation in convolutional neural
  networks,'' in \emph{International conference on artificial neural
  networks}.\hskip 1em plus 0.5em minus 0.4em\relax Springer, 2014, pp.
  281--290.

\bibitem{wei2018deep}
Z.~Wei and X.~Chen, ``Deep-learning schemes for full-wave nonlinear inverse
  scattering problems,'' \emph{IEEE Transactions on Geoscience and Remote
  Sensing}, vol.~57, no.~4, pp. 1849--1860, 2018.

\bibitem{common2002propagation}
L.~T. Common, ``Propagation losses through common building materials 2.4 ghz vs
  5 ghz,'' \emph{E10589, Magis Network, Inc}, 2002.

\bibitem{stavrou2003review}
S.~Stavrou and S.~Saunders, ``Review of constitutive parameters of building
  materials,'' in \emph{Twelfth International Conference on Antennas and
  Propagation, 2003 (ICAP 2003).(Conf. Publ. No. 491)}, vol.~1.\hskip 1em plus
  0.5em minus 0.4em\relax IET, 2003, pp. 211--215.

\bibitem{grosvenor2009time}
C.~A. Grosvenor, R.~T. Johnk, J.~Baker-Jarvis, M.~D. Janezic, and B.~Riddle,
  ``Time-domain free-field measurements of the relative permittivity of
  building materials,'' \emph{IEEE Transactions on Instrumentation and
  Measurement}, vol.~58, no.~7, pp. 2275--2282, 2009.

\bibitem{heaton2008introduction}
J.~Heaton, \emph{Introduction to neural networks with Java}.\hskip 1em plus
  0.5em minus 0.4em\relax Heaton Research, Inc., 2008.

\end{thebibliography}

\appendix
In this appendix, we discuss the experiments that were performed to determine the design parameters of the proposed algorithms. We consider two types of parameters. The first are the radar parameters that include the number and position of the radar antennas with respect to the walls. The second are the GAN network hyper-parameters such as the number of layers, the learning rate and the dropout.
\subsection{Radar Parameters}
\begin{figure}[htbp]
    \centering
    \includegraphics[scale= 0.35]{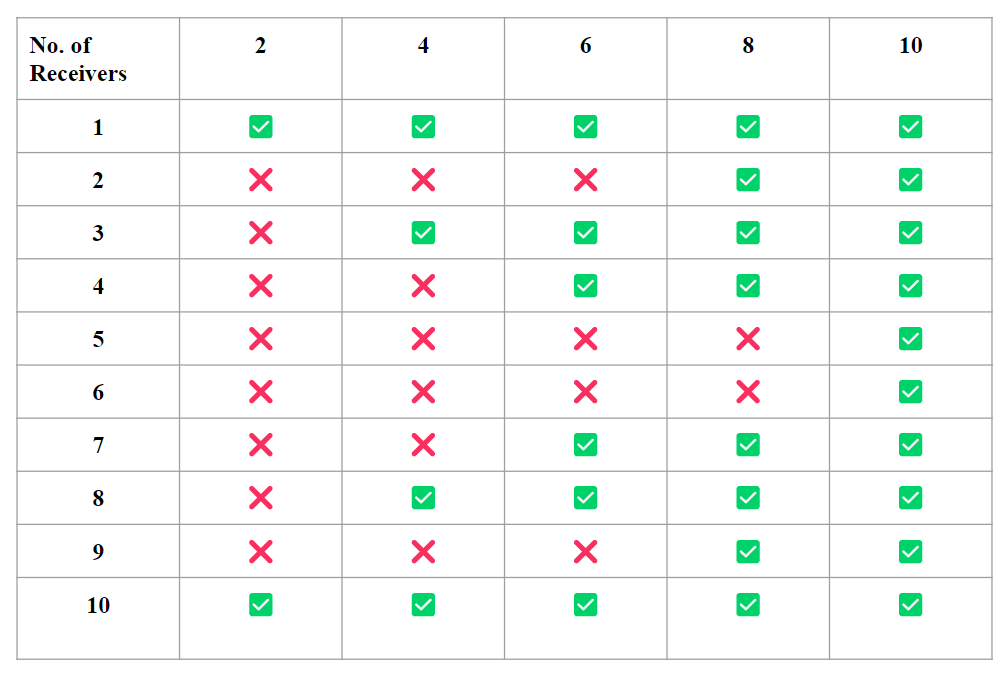}
    \caption{Influence of receiver position on the proposed algorithms.}
\label{fig:Influence_receiver_pos}
    \end{figure}

\textbf{Number of receivers:}
First, we consider the effect of the number and position of the receivers on the performance. We consider the simulation set up shown in Fig.\ref{fig:simulation_setup} where the scattered electric field is gathered at 10 receiver positions that are spaced half wavelength apart. In our experiment, we consider 5 scenarios wherein the data from a subset of the ten receivers are considered in increments of two as shown in Fig.\ref{fig:Influence_receiver_pos}. In other words, in the first scenario, data from receiver 1 and 10 are considered. Then, in scenario 2, data from receivers 1, 3, 8 and 10 are considered and so on. In each case, we ensure that the receivers span the entire aperture space. 
Note that the number of receivers is increased across the five scenarios systematically to include the receivers selected for the previous case.
    \begin{table}[!htbp]
    \centering
    \caption{Influence of number of receivers}   \label{table:Influence_receiver}
    \begin{tabular}{*5c}
    \hline 
    \noalign{\vskip 1pt}
        {Number of receiver}&{GAN-ANNf}&{GAN-ANNt}\\
    \hline 
    \noalign{\vskip 1pt}
     {2}&{0.56}&{0.5}\\
     {4}&{0.54}&{0.51}\\
     {6}&{0.51}&{0.55}\\
     {8}&{0.47}&{0.39}\\
     {10}&{0.23}&{0.25}\\
     \hline 
    \end{tabular}
    \end{table}
The experimental results for the NMSE for each case are presented in Table.\ref{table:Influence_receiver} for GAN-ANNf and GAN-ANNt. Similar trends are observed for GAN-CNNf and GAN-CNNt and hence not reported here. We observe an improvement in the NMSE with the increase in receivers. This may be attributed to increase in information regarding the scattering features within the wall for both GAN frameworks. Further increase in the number of receivers is challenging from a practical deployment standpoint due to requirements of very low inter-element spacing. \\
\textbf{Stand-off position of radar from wall:} Next, we experimented with three different standoff distances of the receivers from the wall - 0.1m, 0.2m, and 0.3m. The resultant NMSE for the  different algorithms are presented in Table.\ref{table:standoff}. 
\begin{table}[!htbp]
\centering
\caption{Influence of standoff position of the receiver from wall.}
\label{table:standoff}
\begin{tabular}{*3c}
\hline 
\noalign{\vskip 1pt}
    {Standoff of receivers}&{GAN-CNNt}&{GAN-CNNf}\\
\hline 
\noalign{\vskip 1pt}
     {0.1m}&{0.17}&{0.2}\\
     {0.2m}&{0.17}&{0.2}\\
     {0.3m}&{0.2}&{0.22}\\
     \hline 
\end{tabular}
\end{table}
The results show that the NMSE changes very slightly across the three cases and hence the performance is not very sensitive to the standoff of the receivers.
\subsection{Network Architecture:}
We experiment with different numbers of layers and dropout rates in the generator and discriminator neural networks. There are some thumb rules towards deciding the network configurations \cite{hinton2006fast,heaton2008introduction}. The total number of hidden nodes is recommended to lie between the size of input and output layer (the number of hidden nodes should be $2/3rd$ the size of the input layer, plus the size of the output layer); the number of hidden nodes should be less than twice the size of the input layer. Following these rules, we experimented with several different configurations for GAN-ANNf as reported in Table.\ref{table:choice_of_architecture}. 
\begin{table}[!htbp]
\centering
\caption{Architecture choices for generator and discriminator for GAN-ANNf}
\label{table:choice_of_architecture}
\begin{tabular}{*4c}
\hline 
\noalign{\vskip 1pt}
    {Generator}&{Discriminator}&{NMSE (0.1)}&{NMSE (0.2)}\\
    \hline 
    {128-256}&{64-32}&{0.56}&{0.57}\\
    {}&{128-64}&{0.51}&{0.52}\\
    {}&{256-128}&{0.46}&{0.35}\\
    {}&{512-256}&{0.55}&{0.31}\\
        \hline
    {64-128}&{64-32}&{0.34}&{0.29}\\
    {}&{128-64}&{0.5}&{0.49}\\
    {}&{256-128}&{0.35}&{0.3}\\
     \hline
    {256-512}&{512-256}&{0.25}&{0.23}\\
    {64-128-512}&{512-256}&{0.5}&{0.36}\\
\hline 
\end{tabular}
\end{table}
The NMSE is studied for different number of layers, hidden nodes per layer and for two different drop-out rates - 0.1 and 0.2.
The exercise is similarly repeated for time domain GAN versions and the results are reported in Table.\ref{table:ANNt_choice_of_architecture}.
\begin{table}[!htbp]
\centering
\caption{Architecture choices for generator and discriminator for GAN-ANNt}
\label{table:ANNt_choice_of_architecture}
\begin{tabular}{*4c}
\hline 
\noalign{\vskip 1pt}
    {Discriminator}&{Generator}&{NMSE (0.2)}\\
    \hline 
    {128-128}&{2048-1024}&{0.43}\\
    {}&{1024-2048}&{0.5}\\
    {}&{512-512-256}&{0.42}\\
    {}&{256-512}&{0.39}\\
    {}&{512-768}&{0.3}\\\hline
    {256-512}&{2048-1024}&{0.49}\\
    {}&{1024-2048}&{0.58}\\
    {}&{512-512-256}&{0.38}\\
    {}&{256-512}&{0.34}\\
    {}&{512-768}&{0.25}\\
\hline 
\end{tabular}
\end{table}
We also tuned two hyper-parameters - the learning rate and the batch size - and report the results in Table.\ref{table:hyper_parameter}.  
\begin{table}[!htbp]
\centering
\caption{Hyper parameter tuning for both the architecture for ANN architecture}
\label{table:hyper_parameter}
{\begin{tabular}{*4c}
\hline 
\noalign{\vskip 1pt}
    {Hyper parameter }&{Value}&{NMSE}\\
    \hline
    {Learning rate}&{0.001}&{0.7}\\
    {}&{0.0002}&{0.23}\\
    {}&{0.0005}&{0.27}\\
    \hline
    {Batch size}&{16}&{0.23}\\
    {}&{32}&{0.24}\\
\hline 
\end{tabular}}
\end{table}
\begin{table}[h]
\centering
\caption{Architecture choices for generator and discriminator for GAN-CNNt}
\label{table:CNNt_choice_of_architecture}
\begin{tabular}{c|cc|ccc|c}
\hline
{} &  \multicolumn{2}{c}{Discriminator} & \multicolumn{3}{c}{Generator} & {}\\
{} &layer1  &layer2  &layer1 &layer2 &layer3  & NMSE \\
\hline
Filters &128 &128    &64  &32  &32  &\\
Kernel &$3\times3$ &$3\times3$    &$5\times5$  &$3\times3$  &$3\times3$  &0.42\\
Stride &$2\times2$ &$2\times2$    &$1\times1$  &$1\times1$  &$2\times2$  &\\\hline
Filters &128 &128    &176  &64  &-  &\\
Kernel &$3\times3$ &$3\times3$    &$5\times5$  &$3\times3$  &-  &0.31\\
Stride &$2\times2$ &$2\times2$    &$1\times1$  &$2\times2$  &-  &\\\hline
Filters &128 &128    &128  &128  &-  &\\
Kernel &$3\times3$ &$3\times3$    &$5\times5$  &$3\times3$  &- &0.17\\
Stride &$2\times2$ &$2\times2$    &$1\times1$  &$2\times2$  &-  &\\\hline
\hline
\end{tabular}
\end{table}
Similarly, the choices for CNN based GAN architectures are reported in Table.\ref{table:CNNf_choice_of_architecture} and Table.\ref{table:CNNt_choice_of_architecture} for frequency domain data and time domain data respectively.
\begin{table}[h]
\centering
\caption{Architecture choices for generator and discriminator for GAN-CNNf}
\label{table:CNNf_choice_of_architecture}
\begin{tabular}{c|cc|ccc|c}
\hline
{} &  \multicolumn{2}{c}{Discriminator} & \multicolumn{3}{c}{Generator} & \\
{} &layer1  &layer2  &layer1 &layer2 &layer3  &NMSE \\
\hline
Filters &128 &128    &64  &32  &32  &\\
Kernel &$3\times3$ &$3\times3$    &$3\times3$  &$3\times3$  &$3\times3$  &0.4\\
Stride &$2\times2$ &$2\times2$    &$2\times2$  &$2\times2$  &$1\times1$  &\\\hline
Filters &128 &128    &176  &64  &-  &\\
Kernel &$3\times3$ &$3\times3$    &$4\times4$  &$4\times4$  &-  &0.38\\
Stride &$2\times2$ &$2\times2$    &$2\times2$  &$2\times2$  &-  &\\\hline
Filters &128 &128    &128  &128  &-  &\\
Kernel &$3\times3$ &$3\times3$    &$4\times4$  &$4\times4$  &-  &0.2\\
Stride &$2\times2$ &$2\times2$    &$2\times2$  &$2\times2$  &-  &-\\
\hline\hline
\end{tabular}
\end{table}
In all of the above cases, the parameters were heuristically selected and the search for the parameters was stopped when we obtained an average NMSE below 0.25.
\end{document}